\documentclass[pra,aps,showpacs,twocolumn]{revtex4}
\usepackage{graphicx}
\begin{document}

\newcommand{\nn}{\nonumber}
\newcommand{\ee}{\end{equation}}
\newcommand{\bea}{\begin{eqnarray}}
\newcommand{\eea}{\end{eqnarray}}
\newcommand{\wee}[2]{\mbox{$\frac{#1}{#2}$}}   
\newcommand{\unit}[1]{\,\mbox{#1}}
\newcommand{\degree}{\mbox{$^{\circ}$}}
\newcommand{\ltish}{\raisebox{-0.4ex}{$\,\stackrel{<}{\scriptstyle\sim}$}}
\newcommand{\vs}{{\em vs\/}}
\newcommand{\bin}[2]{\left(\begin{array}{c} #1 \\ #2\end{array}\right)}
\newcommand{\p}{_{\mbox{\small{p}}}}
\newcommand{\m}{_{\mbox{\small{m}}}}
\newcommand{\tra}{\mbox{Tr}}
\newcommand{\rs}[1]{_{\mbox{\tiny{#1}}}}        
\newcommand{\ru}[1]{^{\mbox{\tiny{#1}}}}

\title{Optical amplifier-powered quantum optical amplification}
\author{John Jeffers}
\affiliation{Department of Physics, Scottish Universities Physics Alliance, University of Strathclyde, John Anderson Building, 107 Rottenrow, Glasgow G4 0NG, UK. john@phys.strath.ac.uk}
\begin{abstract}
I show that an optical amplifier, when combined with photon subtraction, can be used for quantum state amplification, adding noise at a level below the standard minimum. The device could be used to significantly decrease the probability of incorrectly identifying coherent states chosen from a finite set.
\end{abstract}
\maketitle
\section{Introduction}
Deterministic optical amplifiers must add noise photons to any input optical signal in order to satisfy the uncertainty principle \cite{caves}. Typically, amplifiers are based on population inversion, so the amplified signal and noise respectively come from stimulated emission initiated from signal and spontaneously emitted photons \cite{shepherd, glauber, stenholm, yamamoto, jil, qtol}. These two types of output cannot be distinguished, and so the extra noise photons often destroy useful quantum properties of a signal \cite{friberg}, although this does not always render amplifiers useless in quantum optical systems \cite{jj,hamilton1,hamilton2,nha}. 

Perfect deterministic quantum optical amplifiers are proscribed by quantum mechanics. Such devices would be able to transform a coherent state by simply increasing its input coherent amplitude $\alpha$ multiplicatively without adding noise: $| \alpha \rangle \rightarrow |g \alpha \rangle$ with $|g|>1$ as the amplitude gain. The transformation can be performed, but only nondeterministically. The quantum scissors device \cite{peggscissors, babichev} provides a means of perfect amplification in a limited state space \cite{ralph2009,ferreyrol,ralph2010,jjscamp}, and it has been adapted for polarisation qubit amplification \cite{gisin}. Also, the addition and subtraction of photons \cite{marek,fiurasek,zavatta} performs the required transformation with $g=2$ \cite{fiurasek,zavatta}. Weak measurement can also be used as a means of amplification \cite{menzies}. The main drawback with most of these schemes is that they require quantum resources, such as single photons, typically produced in a heralded fashion from downconversion pairs.

The addition/subtraction scheme can also amplify approximately if the single photon addition is replaced by a noisy photon source, where the noise photons have the standard thermal chaotic probability distribution \cite{marek,usuga}
\bea
\label{thermal}
P(n) = \frac{\bar{n}^n}{(\bar{n}+1)^{n+1}},
\eea
and where $\bar{n}$ is the mean photon number. The added noise is displaced in phase space by the input coherent amplitude, after which one or more photons are subtracted from the beam by a weakly reflecting beam splitter and a photodetector. This has the approximate effect of applying the annihilation operator to the state. The associated bosonic enhancement biases the output towards higher photon numbers, and thus provides the amplification. 

This noise-powered amplification (NPA) mechanism produces an amplifier that works better than the deterministic limit. Although the amplification is not perfect, it is good enough to reduce the phase variance of an amplified coherent state to significantly below its initial value. This would help, for example, in distinguishing two low amplitude coherent states, as is required in some quantum communication schemes \cite{andersson,hamilton1}. Successive photon subtractions cause the state to spread asymmetrically in phase space, but not enough to offset the increase in amplitude caused by the boson enhancement. 

There are at least two other advantages to this system over the earlier schemes. Not only does it remove the state space limitation for the amplifier, but more significantly it shows that better than deterministic amplification can be performed {\it without} the use of quantum resources such as single photons. 

As has already been noted, the output of standard deterministic optical amplifiers has two components - perfectly amplified input, and added noise \cite{shepherd}. For a coherent input state $| \alpha \rangle$ the output of the optimal optical amplifier with intensity gain $G$ consists of photons corresponding to a perfectly amplified state $| \sqrt{G} \alpha \rangle$, and a thermal distribution (\ref{thermal}) with a minimum mean number of photons $\bar{n}=G-1$. The mean number of added photons is larger if the amplification medium is not perfectly inverted. However, there has been significant progress in quantum optical amplification, and the noise limit has been closely approached \cite{josse,pooser}. 

\begin{figure}[h]
\centering
\includegraphics[height=4cm]{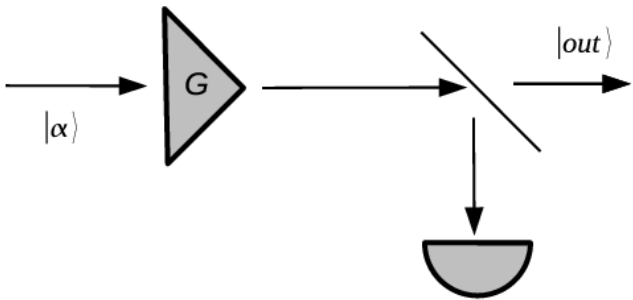}
\caption{Amplification scheme: Deterministic amplification is followed by photon subtraction. In Refs. \cite{marek,usuga} the amplifier was replaced by the addition of noise.}
\label{fig1}
\end{figure}

The two components of the amplifier output have different sources, and so (apart from both being a function of $G$) they are independent of one another. Thus it is already clear that amplifiers add noise in exactly the same way as in the original NPA scheme \cite{marek,usuga}, a fact that was already pointed out in \cite{usuga}. This second scheme is here called amplifier-powered amplification (APA). The combination of amplification (together with the accompanying addition of phase-insensitive noise) and photon subtraction (Fig. \ref{fig1}) provides a significant reduction in phase variance for the output of the amplifier. This will provide a commensurate improvement in identifying different coherent states with the same magnitude. 

\section{Amplifier output and gain}
The density operator for the output of a perfectly-inverted amplifier can be written as \cite{shepherd,jedrkiewicz}
\bea
\nn \rho\ru{out}_{n,m} &=& \sqrt{\frac{(G-1)^{n+m}}{G^{n+m+2}}}\sum_{p=0}^n \sqrt{\left( \begin{array}{c} n \\ p \end{array}\right)\left( \begin{array}{c} m \\ m-n+p \end{array}\right)}\\
&\times& \frac{\rho\ru{in}_{p,m-n+p}}{(G-1)^{p+(m-n)/2}},
\eea
where $\rho\ru{in}_{p,q}$ are the density matrix elements of the input state, which for the coherent state considered are
\bea
\rho\ru{in}_{p,q} = e^{-|\alpha|^2} \frac{\alpha^p \alpha^{*q}}{\sqrt{p!q!}}.
\eea
Successive photon subtractions transform the density matrix elements according to
\bea
\label{subtract}
\rho^M_{n,m} = \frac{1}{\mathcal{N}} \sqrt{(n+1)(m+1)}\rho^{M-1}_{n+1,m+1},
\eea
where $M$ is the number of subtracted photons and $\mathcal{N}$ is chosen so as to normalize the state. 

\begin{figure}[h]
\centering
\includegraphics[height=5.5cm]{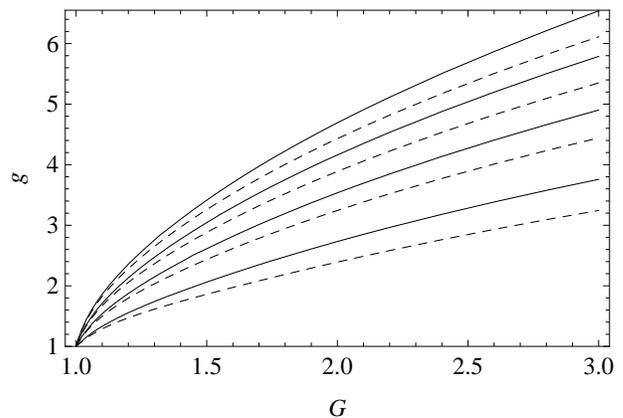}
\caption{Device amplitude gain $g$ as a function of amplifier intensity gain $G$ for the APA (solid lines) and as a function of the number of added photons $G-1$ for the NPA (dashed lines). For each set of curves the number of subtracted photons ranges from 1 (bottom line) to 4 (top line).}
\label{fig2}
\end{figure}

In order to characterize the states produced by the APA the gains of the two devices are first examined. The coherent state with maximum overlap with the device output state forms the nominal output state for the device $|g\alpha \rangle$. For the APA $g$ is a function of the amplifier gain $G$ whereas for the NPA it is a function only of the the number of added photons.  In Fig. \ref{fig2} $g$ is plotted for both devices for an input coherent amplitude $\alpha=0.5$ corresponding to a mean coherent input photon number of 0.25, similar to that used in reference \cite{usuga}. 
For comparison purposes it is assumed here that the mean number of added photons in the noise-based bystem is the same as the mean number of noise photons added by the amplifier, $G-1$. For both devices larger gains are accessed for more subtracted photons. However, the gain of the APA for a given number of subtracted photons is larger than the corresponding gain for the NPA. As a simple example for $G=2$ and thus $\bar{n}=1$ the amplitude gains of the two devices are $g=$2.73 and 2.39 corresponding to intensity gains of about 7.5 and 5.7. Thus there is a straightforward advantage in using an amplifier rather than simply adding noise. Note that the gains of both devices depend upon the input photon number. Both gains will be larger for lower amplitude input coherent states. The photon subtraction effect in Eq. (\ref{subtract}) does not bias distributions of higher photon numbers as much as it does those of lower photon numbers, and so the gains of both devices will decrease with input coherent amplitude.

\section{Output fidelity and phase noise}
\begin{figure}[h]
\centering
\includegraphics[height=5.5cm]{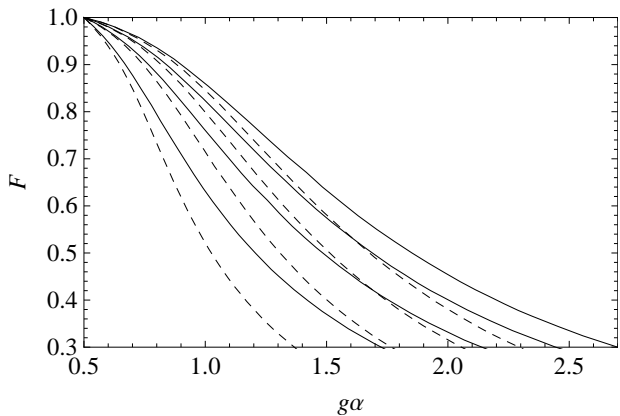}
\caption{Device fidelity as a function of output nominal coherent amplitude for the APA (solid lines) and the NPA (dashed lines). For each set of curves the number of subtracted photons ranges from 1 (bottom line) to 4 (top line).}
\label{fig3}
\end{figure}

Gain is not the most important feature for quantum amplifiers; noise is crucial. This becomes apparent when one considers the fidelity of the output state produced when compared with the nominal output state $|g\alpha \rangle$, 
\bea
F=\tra\left( \hat{\rho} |g\alpha \rangle \langle g\alpha| \right).
\eea
Fidelity does not give full information about the phase-space distribution of the state, which will not have circular Wigner function contours \cite{marek}. However, it does give a reasonable indication of the quality of amplification. Therefore the fidelity is plotted as a function of nominal output coherent amplitude in Fig. \ref{fig3}. 
Higher fidelities are achievable for a particular desired nominal state if the number of subtractions is increased and a consequent lower value of $G$ (APA) or $\bar{n}$ (NPA) is used. However, for the NPA, higher values of $g\alpha$ are only achievable at a significantly lower fidelity than the APA for the same number of subtracted photons. The obvious reason for this is that a larger part of the state is already coherent due to the APA amplifier action. Photon subtraction has no effect on this component, but it is not required to do so.

\begin{figure}[h]
\centering
\includegraphics[height=5.5cm]{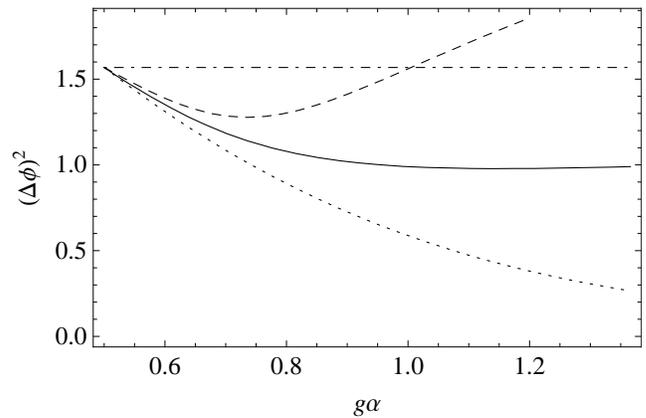}
\caption{Phase variance as a function of output coherent amplitude for the APA (solid line) and the NPA (dashed line) for one subtracted photon. The dotted line corresponds to the phase variance of a pure coherent state, and the dot-dashed line to the phase variance of the initial state.}
\label{fig4}
\end{figure}

\begin{figure}[h]
\centering
\includegraphics[height=5.5cm]{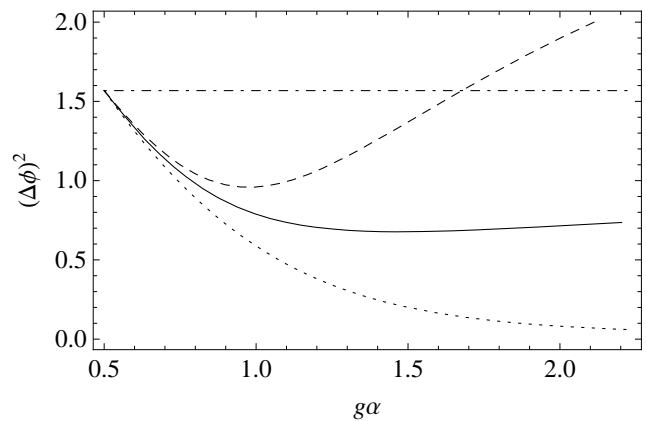}
\caption{As Fig. 4, but for two subtracted photons.}
\label{fig5}
\end{figure}

The higher fidelity states produced by the APA ought to be apparent in the phase variance of the output \cite{barnettradmore}. This is most straightforwardly calculated for an input state of zero mean phase (e.g. for real positive $\alpha$), and in a phase window of $(-\pi, \pi]$. Then the phase variance of a density operator with number state elements $\rho_{n,m}$ can be written as
\bea
\left( \Delta \phi \right)^2 = \sum_{n,m=0}^\infty \left[ \frac{\pi^2}{3} \delta_{nm} + \left( 1- \delta_{nm} \right) \frac{(-1)^{m-n}}{(m-n)^2}\right] \rho_{n,m}.
\eea
A deterministic amplifier cannot decrease this quantity, and even an optimal deterministic amplifier increases the phase variance slightly. 
The variance is plotted in Figs. \ref{fig4} and \ref{fig5} as a function of the nominal output coherent amplitude for both the APA and the NPA systems, for one and two subtracted photons in order to stay close to what is at present experimentally achievable. Also shown is the phase variance of the nominal coherent state $|g\alpha \rangle$. Here the full advantage of using an amplifier shows itself. The APA outperforms the NPA across the whole domain of nominal coherent amplitudes. Although neither system approaches the phase variance of the nominal coherent state, the minimum variance achievable is significantly lower for the APA. For one subtracted photon the APA minimum variance is 75\%, and for two it is 70\%, of the minimum NPA values. Furthermore, the variance minimum is a flat function, so a large range of amplifier gains will still produce a significantly lower variance. 

The improvement is apparent even for relatively low gains, which at first might seem slightly mystifying, as the extra coherent component provided by the output of the amplifier is small. The reason for such a large difference is a combination of two factors. First the  $G$-gain enhancement for the APA seen in Fig. \ref{fig2} means that a greater number of photons must be added by the NPA to compensate. As the optimal mean number of added photons required to minimize the phase variance of the NPA is quite low (in \cite{marek} for $\alpha=0.48$ the minimum phase variance occurs for $\bar{n}\simeq 0.25$), the upturn in the phase variance occurs at low values of $g$. Second, the fidelity difference between the two systems displayed in Fig. \ref{fig3} shows that there is a much larger coherent component of the APA output state, so its variance will follow that of the nominal output state more closely at higher gains. Note that the phase variance of a pure coherent state drops dramatically with coherent amplitude. Thus even a relatively small increase in the coherent component of the output state can produce a large difference in phase variance. This increase in the coherent component is exactly what the amplifier provides. For higher output nominal coherent amplitudes the APA advantage due to amplification of the input coherent state is greater, but the photon subtraction effect is smaller. A combination of these two effects is responsible for the flatness of the APA phase variance as a function of gain. This flatness is clearly not possible for the NPA. 

A simple use for the type of amplifier described here might be to distinguish different weak coherent states, say $|\alpha\rangle$ and $|-\alpha \rangle$. These two states are not orthogonal, with squared overlap $e^{-2|\alpha|^2}$. Use of the composite amplifier and photon subtractor (or use of the NPA) will reduce this overlap, and increase the orthogonality of the two states. This is quantified here using the mixed state formula for fidelity \cite{jozsa}, 
\bea
F=\left[\tra \sqrt{\sqrt{\hat{\rho}(\alpha)}\hat{\rho}(-\alpha)\sqrt{\hat{\rho}(\alpha)}}\right]^2,
\eea
which gives the probability that the amplified state $\hat{\rho}(\alpha)$ would be incorrectly identified as $\hat{\rho}(-\alpha)$ in a yes/no measurement test. For $\alpha=0.5$ the lowest value that can be obtained with the APA is $F=0.178$ for one subtracted photon and $F=0.075$ for two. These figures represent about 50\% and 20\% of the pre-amplification value of 0.368. The APA dramatically increases the orthogonality of the original coherent states.

It is also worth noting the effect of varying the input coherent amplitude on the results presented here. The bosonic enhancement effect works best on low photon numbers, so although the overall phase variance is reduced most for a higher amplitude input coherent state, the decrease in variance will be much more pronounced for low amplitude states. The flatness of the phase variance as a function of nominal output coherent amplitude does not vary rapidly for different input coherent amplitudes. This means that setting a particular intensity gain $G$ for the amplifier will provide an output state with near minimum phase variance for a large range of input coherent amplitudes. This is in contrast to the NPA, where the more pronounced phase variance minima dictate a closer tailoring of the added thermal noise to the input coherent state.

\section{Conclusions}
In conclusion, it has been shown that the use of an optimal amplifier combined with photon subtraction produces an output state, from an input coherent state, with a phase variance which is greatly decreased from its initial value; the variance is also significantly below the value possible for thermal noise addition and photon subtraction. For such an optimal amplifier, which adds the minimum number of noise photons, the gain $G$ has the maximum possible effect on the input state. Typically population inversion amplifiers are not optimal, as they are not 100\% inverted, but there has been a significant progress in reaching optimality in feed-forward \cite{josse} and four-wave mixing \cite{pooser} systems, so it is not unrealistic to assume this. Furthermore, Raman amplifiers can also approach optimality over a large range of gains \cite{battle,swanson}. The effect of a non-optimal amplifier would be simply to produce results between those of the optimal APA and the NPA. The device gain and fidelity would be reduced, and the phase variance increased from the optimal APA values. There would always be some advantage to using an amplifier, no matter how poor, over noise addition as even the noisiest amplifier will produce some amplified coherent output.

The experimental feasibility of the APA system depends on photon subtraction and amplifier quality. Photon subtraction has been used in several experiments and is based on postselection using a highly-transmitting beam splitter and a photodetector. For near-ideal subtraction the reflection coefficient should be close to zero, but this reduces the likelihood of subtraction success. This effect hs been overlooked here, but it was considered in \cite{marek}, where it was shown not to be a significant problem for the observation of the variance reduction. This was borne out by experiment \cite{usuga}.

The utility of amplifiers in quantum optics has traditionally been limited by the necessary addition of noise. Systems which add noise and then use photon subtraction can at least partially circumvent this limitation. If the noise addition system is an amplifier itself, even one with very low gain, then a small fraction of the added photons are coherent. This leads to a much improved device, which could with present technology be used to distinguish between such non-orthogonal states with little ambiguity. Furthermore, it suggests that low noise amplification systems will have future uses in quantum communication schemes.

The author would like to acknowledge the UK Engineering and Physical Sciences Research Council for financial Support.

\end{document}